\documentclass[twocolumn]{aastex63}

\received{2021}
\revised{2022}
\accepted{2022}

\submitjournal{ApJ}

\usepackage{graphicx}	
\usepackage{amsmath}	

\newcommand{\degree}{$^{\circ}$}

\shorttitle{GRB170202A}
\shortauthors{Gendre et al.}

\begin{document}



\title{Modeling GRB 170202A fireball from continuous observations with the Zadko and the Virgin Island Robotic Telescopes}

\correspondingauthor{Bruce Gendre}
\email{bruce.gendre@uwa.edu.au}

\author[0000-0002-9077-2025]{B. Gendre}
\affiliation{Etelman Observatory Research Center, University of the Virgin Islands, St. Thomas, USVI}
\affiliation{OzGrav-UWA, 35 Stirling Highway, M013, 6009 Crawley, WA, Australia}

\author[0000-0001-6866-1436]{N. B. Orange}
\affiliation{Etelman Observatory Research Center, University of the Virgin Islands, St. Thomas, USVI}
\affiliation{OrangeWave Innovative Science, LLC, Moncks Corner, SC 29461, USA}

\author{E. Moore}
\affiliation{OzGrav-UWA, 35 Stirling Highway, M013, 6009 Crawley, WA, Australia}

\author{A. Klotz}
\affiliation{IRAP, Université Toulouse III Paul Sabatier, 9, avenue du Colonel Roche, BP 44346, 31028 Toulouse Cedex 4, France}

\author[0000-0002-7795-9354]{D. M. Coward}
\affiliation{OzGrav-UWA, 35 Stirling Highway, M013, 6009 Crawley, WA, Australia}

\author{T. Giblin}
\affiliation{United States Air Force Academy, Colorado Springs, CO 80840, USA}

\author{P. Gokuldass}
\affiliation{Etelman Observatory Research Center, University of the Virgin Islands, St. Thomas, USVI}
\affiliation{Department of Chemical and Physical Sciences, University of the Virgin Islands, St Thomas, USVI}

\author{D. Morris}
\affiliation{Etelman Observatory Research Center, University of the Virgin Islands, St. Thomas, USVI}
\affiliation{Department of Chemical and Physical Sciences, University of the Virgin Islands, St Thomas, USVI}

\begin{abstract}
We present coordinated observations of GRB 170202A carried out by the Zadko and the Virgin Island Robotic Telescopes. The observations started 59\,s after the event trigger, and provided nearly continuous coverage for two days due to the unique location of these telescopes. We clearly detected an early rise in optical emission, followed by late optical flares. By complementing these data with archival observations, we show GRB 170202A is well described by the standard fireball model if multiple reverse shocks are taken into account. Its fireball is evidenced to expand within a constant density interstellar medium, with most burst parameters consistent with the usual ranges found in literature. The electron and magnetic energy parameters ($\epsilon_e, \epsilon_B$) are orders of magnitude smaller than commonly assumed values. We argue that the global fit of the fireball model achieved by our study should be possible for any burst, pending the availability of a sufficiently comprehensive data set. This conclusion emphasizes the crucial importance of coordinated observation campaigns of GRBs, such as the one central to this work, to answer outstanding questions about the underlying physics driving these phenomena.

\end{abstract}


\section{Introduction}\label{Introduction}

Gamma-Ray Bursts (GRBs) are bursts of high energy photons coming from the far edge of the Universe \citep{kle73,met97}. They are usually considered as ideal laboratories for studying extreme physics in the Universe \citep[see][for a review]{zha18}. To do so, a standard unified model, the Fireball Model \citep{ree92, mes97, pan98} has been patiently constructed and validated against numerous observations \citep[see e.g.][]{klo09a,klo09b,ved09}. With this model and a subset of ground and space-based observations of a GRB event, it is possible to gain access to the surrounding medium of the progenitor, the physical conditions inside the plasma shell producing the event, and the geometrical constraints of the system. A prime example of this is the study of GRB 110205A by \citet{gen12}.

Such a straightforward study of GRBs is unfortunately not very common. This stems, in part, from difficulties with the fireball model and the large degrees of freedom to explain observed features \citep{gen09}. For instance, technically, all forms of a surrounding medium density (from a constant one to a highly chaotic one) can be accommodated by the model \citep{pan00}. In practice, either a constant interstellar medium (ISM) or wind environment is considered sufficient to represent the full-extent of reality \citep[e.g.][]{yos03}. These practices introduce inconsistencies between observed data and the use of {\it ad hoc} model adjustments, and leave unexplored questions of whether or not more complex environments could explain the observations \citep{kum15}. The same can be said for any of the free parameters of the fireball model, which must be fully described to investigate key questions related to the extreme physics of GRBs. For instance, the magnetic equipartition parameter of the fireball model, $\epsilon_B$, quantifies the energy available from magnetic fields. However more information is needed to fully deduce the complexity and nature of the magnetic field of the fireball itself \citep[see][and references within]{zha11}, especially when considering the fact that these fields can be self-generated by the fireball or be present as ``primordial" magnetic fields \citep{zha03}.

The standard fireball model parameters are more easily constrained when the peak of the afterglow emission is observed at a given frequency. Observations that capture the early rise in optical emission are therefore a key to constraining GRB models, again akin to the study of GRB 110205A by \citet{gen12}. In this paper we carry out a similar in-depth study of GRB 170202A, an event also accompanied by an extended period of rising optical emission.

Our study focuses on results from the Zadko Telescope and Virgin Islands Robotic Telescope (VIRT) optical observations, presented in Section \ref{telescopes}. The data analysis is explained in Section \ref{analysis}, followed by a discussion about the observations in the context of the fireball model in Section \ref{fireball}. In Section \ref{discu}, we present a more broad discussion about the constraints we placed on the fireball model, before concluding. For our analysis we adopt a standard $\Lambda$CDM cosmology model with $H_0 = 70$\,km$^{-1}$s\,Mpc$^{-1}$, $\Omega_{M} = 0.27$, and $\Omega_{\Lambda} = 0.73$. All errors are quoted at the 90\% confidence level, and we use the standard notation for a given parameter P, $P_x = P \times 10^x$ (when no indices are given, a value of $x = 0$ is assumed).

\section{Observations}\label{telescopes}

\begin{table*}
	\centering
	\caption{Photometric data  of GRB 170202A extracted from GCN circulars and used in our analysis. For each data point we provide the start and end time, filter, and 90\% confidence level. All magnitudes are expressed in the AB system, and have been converted from the GCN where needed. The data has not been corrected for Galactic or host extinction.}
	\label{table_data_gcnc}
	
	\begin{tabular}{cccccc}
	\hline\hline
	Start (s) & End (s) & Filter & mag & 2$\cdot\sigma$ & Reference \\
	\hline
	39  & 69 & Rc & $>$17.88 & --   & \citet{sai17} \\
	77  & 107    & Rc & 17.19    & 0.12 & \citet{sai17}\\
	116    & 146    & Rc & 16.56    & 0.09 & \citet{sai17}\\
	320    & 350    & R & 16.53    & 0.30 & \citet{mor17} \\
	6150   & 6450   & R & 19.16    & 0.07 & \citet{son17} \\
	10200  & 10680  & R & 19.65    & 0.15 & \citet{gui17a} \\ 
	15480  & 15480  & r & 20.10    & 0.10 & \citet{kru17} \\
	57504 & 58044 & r & 21.10 & 0.08 & \citet{im_17a} \\
	57717  & 57957  & R & 21.10    & 0.15 & \citet{gui17b} \\
	141470 & 142010 & r & 21.8 & 0.15 & \citet{im_17b} \\ 
	\hline
	\end{tabular}

\end{table*}

\subsection{High energy observatories}

The Burst Alert Telescope \citep[BAT,][]{bar05} onboard the \textit{Neil Gehrels Swift Observatory} \citep{geh04} triggered on GRB 170202A (trigger=736407) at 18:28:02 UT 2nd February 2017 \citep{rac17}. The duration of the event was $T_{90}$ (15-350 keV) = 46$\pm$12~s \citep{bar17}. The XRT began observing the field 72.5\,s after the BAT trigger \citep{dav17} while UVOT started its observations 83\,s after the BAT trigger \citep{kui17}. Both instruments identified a candidate afterglow at coordinates R.A. = 10:10:03.49 and Declination = +05:00:41.8 \citep{osb17}.

Konus-Wind also triggered on this event, and found the prompt spectrum was best fit by a power law with exponential cutoff, with $E_{peak} = 247^{+166}_{-86}$ keV and a fluence $S = (5.9 \pm 1.4)\times10^{-6}$ erg cm$^{-2}$ in the 20-10,000 keV range \citep{fre17}.

The redshift of this event was measured  using OSIRIS at the 10.4m GTC telescope at the Roque de los Muchachos observatory (La Palma, Spain). The observation started 4.91 hr after the burst, and derived a redshift of 3.645 \citep{pos17}. This translates to an isotropic energy of $E_{iso}$ $\sim1.7\times10^{53}$ erg and $E_{p,i} =  1150$ keV, making this burst compatible with the Amati relation \citep{ama06} when taking into account intrinsic variability of the relation. Assuming a 30\% radiative efficiency, this leads to a total energy budget of $E_0 = 5.67 \times 10^{53}$ ergs.

\subsection{The Zadko Telescope}

The Zadko Telescope \citep{cow10,cow17} is a 1 m f/4 Ritchey-Chretien telescope situated in the state of Western Australia at longitude, 115\degree 42' 49'' E, latitude, 31\degree 21' 24'' S. The telescope was fitted at the time of GRB 170202A observations with an Andor camera with a back-illuminated CCD covering 27'\,$\times$\,27' of field of view, and several filters (SLOAN g’, r’, i’, Clear). The telescope is fully robotized, and specialized in transient source astronomy with two observation modes: a routine mode, following a schedule built every 6 hours; and an alert mode responding in less than 10 seconds to any new transient alert, bypassing the schedule and starting unplanned observations.

In both alert and routine modes, all interactions with the telescope are made remotely via internet. Observation requests can be scheduled at any time up to 15 minutes before the start of the night for the routine observations, and at any time for the alert system. The scheduling format is simple, using a text file for the routine mode and a VO-Event packet for the alert mode.

Zadko began observations of GRB 170202A 59s after the BAT trigger, and followed the event for two consecutive nights. The observations have been reported in \citet{klo17a} and \citet{klo17b}.

\subsection{The Virgin Island Robotic Telescope}
The Etelman Observatory is a Research Center of the University of the Virgin Islands \cite[UVI;][]{ora21}, located at longitude, 64\degree 57' 24'' W, latitude, 18\degree 21' 09'' N \citep{nef04}. It is home to a Torus, Inc. robotic system coined the Virgin Islands Robotic Telescope \citep[VIRT,][]{gib04,mor18}. The VIRT is a 0.5 m f/10 Cassegrain telescope on an equatorial mount fork. The telescope instrumentation includes a 12-position Finger Lakes Instrumentation (FLI) filter wheel (presently housing standard Johnson-Cousins UBVRI and Clear filters) and a FLI ProLine 4240 CCD camera containing a back-illuminated Marconi 42-40, 2048\,$\times$\,2048 pixel sensor. The 13.5 micron pixels provide a 20'\,$\times$\,20' field of view and spatial sampling of 0.5\,arcsec pixel$^{-1}$. Its full-frame readout time at full resolution is approximately 2s. 

VIRT is fully robotized, and can be operated both on-site and remotely through queue-scheduled, and direct user orders. Many nights are characterized by rapidly changing weather conditions (frequent, but usually brief, precipitation events lasting less than 30 minutes), and automated weather related shutdown capabilities remain in-development. As such, an on-site observer oversees the VIRT observing program. The supporting infrastructure of VIRT monitors, and warns observers to transient events, with options to bypass queued-scheduled observing to start unplanned observations \citep{ora21}. Observations for the VIRT can be scheduled through a web interface, which are compiled approximately every four hours. 

The VIRT took exposures of the GRB 170202A for two consecutive nights; however data from the second night were lost during Hurricanes Irma and Maria that struck the facility later that year \citep{gen19}. The first night of observation has been reported in \citet{gen17}.

\subsection{Other Instruments}

Several other instruments responded to the {\it Swift} alert, and published their results in various GCNs. We retrieved only those data taken in red bands with a precise date of the observation. We then converted all magnitudes expressed in the Vega system to obtain a homogeneous sample of AB magnitudes. These data are presented in Table \ref{table_data_gcnc}. Note that they have not been corrected for effects {\bf of} extinction. For GRB 170202A, the Galactic extinction is negligible \citep[E$_{(B-V)}$ = 0.02 mag, A$_R$ = 0.06 mag,][]{sch11}.

GRB 170202A occurred while the Middle-East and Asian regions were at night, and thus most of the earlier observations are reported from Asia. The diameter range of the involved instruments (30\,cm to 2.2\,m), and its mean of about 1\,m provided deep observations. However, given the redshift of this event, they are not deep enough to cover the entire relativistic part of the optical light curve; specifically, observations of the putative jet break in the optical band are missing (see below). The bulk of observations were done in the clear or red filter, with some points in the infrared. Again, we checked that the data used for our study were homogeneous in terms of observation, i.e., similar filters, and that each was reported absorbed (i.e. raw).

Lastly, radio observations were performed by AMI, but no detection was reported \citep{moo17}. We incorporated the upper limit derived from these observations into our study as it additionally constrained our data modeling.

\section{Data reduction and spectral analysis}
\label{analysis}

\subsection{The Swift/XRT data} 

We retrieved the {\it Swift} XRT data available at the main HEASARC archive, and reprocessed them using the latest version of the Ftools and CALDB (i.e. 6.26.1 and 20190910, respectively) available at the time of processing. The journal of the observations are provided in Table \ref{table_swift}.

The XRT started observations 92\,s after the trigger in Window Timing (WT) mode. The instrument switched to the Photon Counting (PC) mode 225\,s after the trigger. Between that time and 4990\,s after the trigger, the PC data suffered from moderate pile-up, and we used the standard method of \citet{vau06} and \citet{rom06} to compare the observed PSF and the predicted one. We found that removing the inner 12 pixels solved the pile-up issue, and for the first observation in PC mode we used an annulus shape to extract the light curve and spectrum. The successive observations were processed using a standard circular region of radius 25\,arcseconds.

\begin{figure}
	\centering
		\includegraphics[width=9.8cm, bb=90 280 550 580]{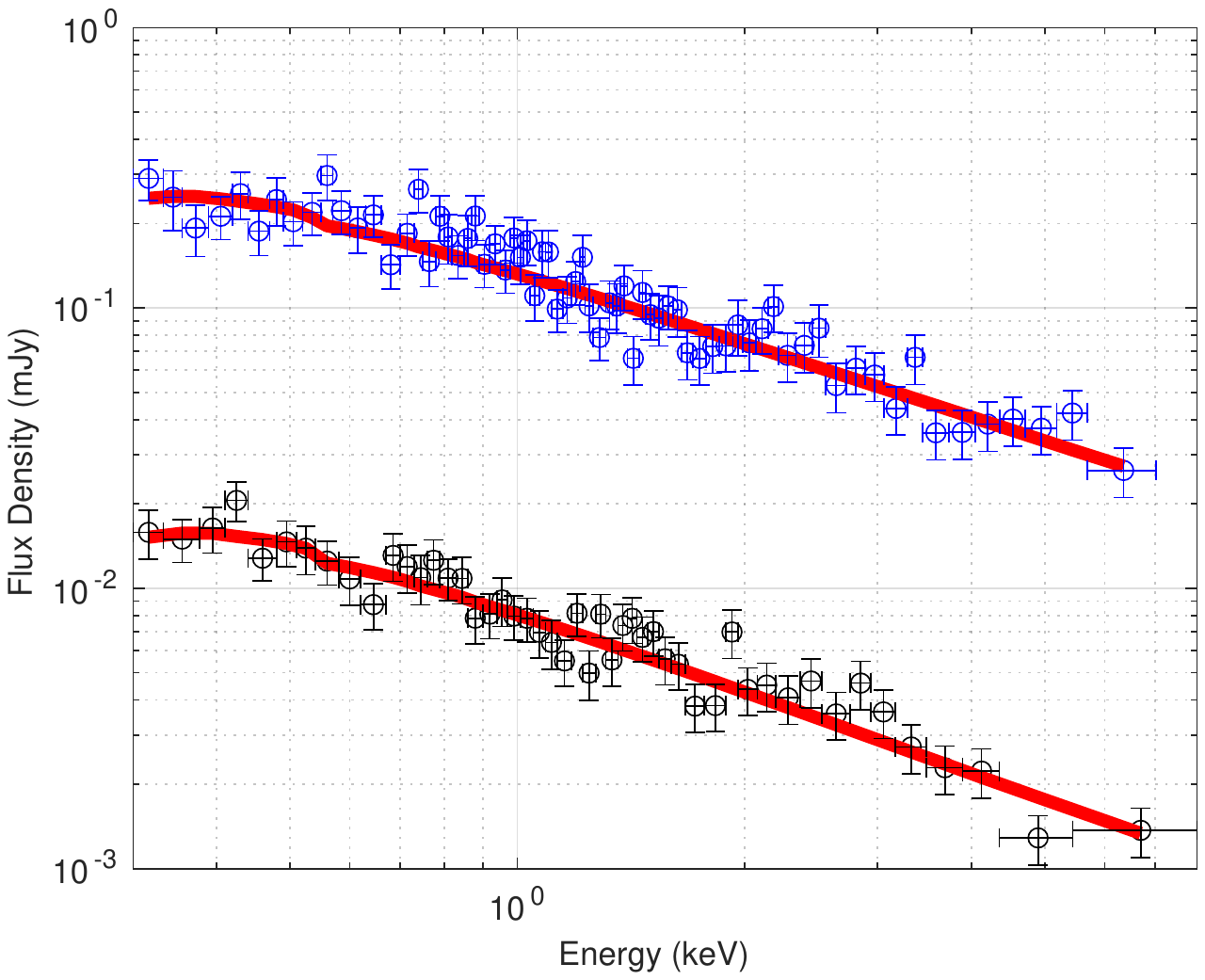}
		\caption{Unfolded X-ray spectrum of GRB 170202A. Top, the WT data; bottom, the PC data taken during the segment number 0 of the observation. The red lines represent the best fit model. No obvious feature is visible in the data. \label{spec_x}}
\end{figure}

In order to reduce uncertainties, we used only the non-piled up data for spectral analysis. The spectra were fit with XSPEC using a simple power law absorbed twice by our galaxy and the host galaxy. The WT and PC spectra are consistent with the same spectral parameters. They are listed in Table \ref{table_specX}. As seen in these results, an upper limit was obtained for the host extinction that is attributable to the large redshift of GRB 170202A. The overall fit to the spectra was good with no evidence of extra-features (see Fig. \ref{spec_x}).

\begin{figure*}
	\centering
		\includegraphics{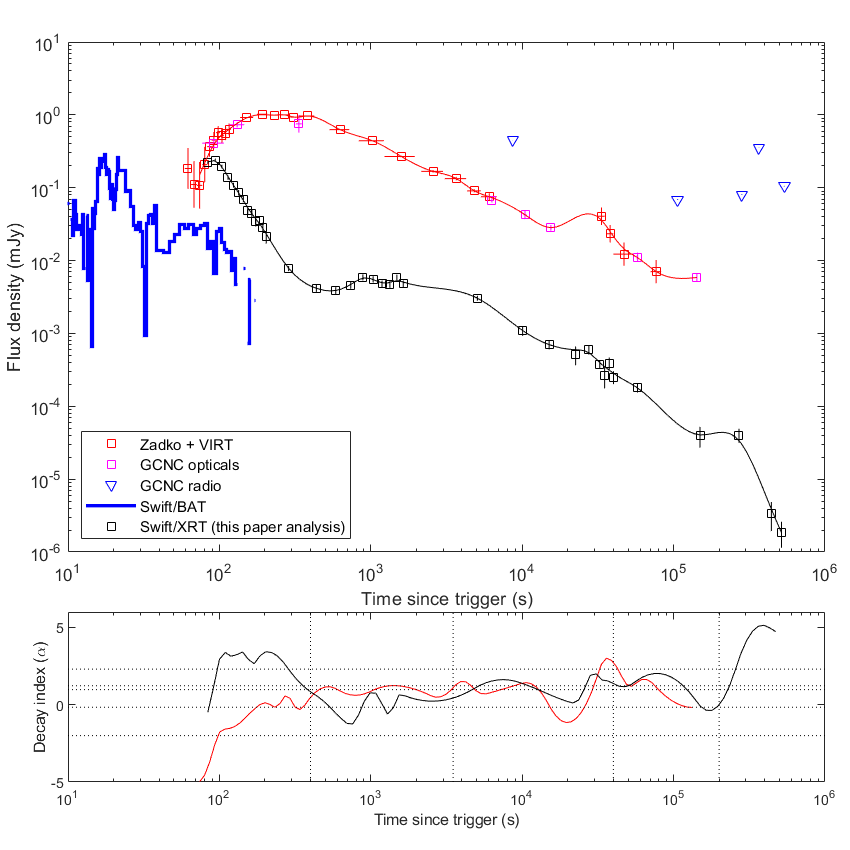}
		\caption{\label{fig_lc_all}Top: Light curves of GRB 170202A in $\gamma$, X-rays, optical R and radio bands. Note that the X-ray light curve is expressed at 1 keV band and unabsorbed. Bottom: Spline fit to the light curve data using a small smoothing factor to detect correlated flares between the various bands. The horizontal dotted lines are decay indices expected from the various temporal segments of the fireball model ($-2, -1/6, (3p-2)/4, (3p-3)/4,$ and $p$).}
\end{figure*}

We extracted three light curves: a soft band (0.5-2.0 keV), a hard band (2.0-10.0 keV), and a large band (0.3-10.0 keV), in order to perform a study of the hardness ratio. We did not find any variability within the error bars, indicating our initial spectral model was valid for the entirety of observations analyzed. We then converted the count rate light curves into energy (flux density or flux, depending on needs) using our spectral model. The resultant data {\bf are} presented in Fig. \ref{fig_lc_all}

\begin{table}
	\centering
	\caption{Journal of the {\it Swift} observations.}
	\label{table_swift}
	\begin{tabular}{cccc}
	\hline\hline
	Segment & Start (s) & End (s) & Comment \\
	\hline
	0 &     86 &   1841 & Pile-up in PC mode\\
	1 &   4990 &  41730 & --\\
	2 &  57687 &  59667 & -- \\
	3 & 149492 & 151480 & --\\
	5 & 207123 & 335071 & --\\
	6 & 436613 & 449494 &  --\\
	7 & 511279 & 522960 & --\\
	8 & 574342 & 574774 & No detection\\
	9 & 626096 & 632254 & No detection\\
	\hline
	\end{tabular}
\end{table}

\begin{table}
	\centering
	\caption{Spectral parameters of the X-ray afterglow  of GRB 170202A. We only performed a fit on the WT mode (from 92--225 s) for better statistics, as the PC mode suffered from pile-up issues. We obtained a $\chi^2_\nu$ value of 0.92 for 82 degrees of freedom.}
	\label{table_specX}
	\begin{tabular}{ccc}
	\hline\hline
	Parameter & Value  & Comment \\
	\hline
	Galactic N$_h$ (cm$^{-2}$)      & $2.30 \times 10^{20}$ & Fixed\\
	Redshift                        & 3.645                 & Fixed\\
	Extragalactic N$_h$ (cm$^{-2}$) & $< 1 \times 10^{22}$  & --\\
	Spectral index $\beta$          & 0.94  $\pm 0.09$      & --\\
	\hline
	\end{tabular}
\end{table}

The 2-10 keV band fluxes are found to be $2.7 \times 10^{-10}$\,ergs cm$^{-2}$ s$^{-1}$ and $1.6 \times 10^{-11}$\,ergs cm$^{-2}$ s$^{-1}$ for the WT and PC mode, respectively.

\subsection{The Zadko data}

\begin{figure}
	\centering
		\includegraphics[width=8.5cm]{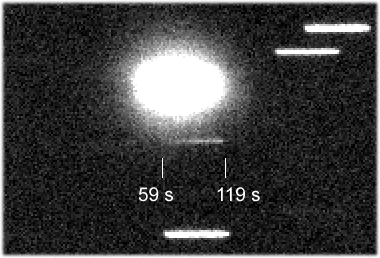}
		\caption{Trailed image of GRB 170202A taken by Zadko. The rising part of its afterglow is clearly observed. \label{fig_trail}}
\end{figure}

The Zadko telescope started observations 59\,s after the BAT trigger using the trailing mode observation described in \citet{klo06}. Through this technique, stars appear as line segments of constant intensity, and GRB afterglows appear as a line segment with intensity fluctuations that follow its flux variations.
Fig.~\ref{fig_trail} presents the trailed image of GRB 170202A from Zadko. Note that its drift length was specifically adopted to cover 30 pixels (40\,arcsec), with a thickness of 2.8\,arcsec.

The afterglow of GRB 170202A was located close to a star of 10th magnitude whose diffuse light overlapped its afterglow trail (see Figure~\ref{fig_trail}). To extract the profile of the afterglow trail we first subtracted a symmetrical component, with respect to the center of the bright star in the trailed image, to obtain a flat background around it. 
The resultant profile of the trail was then compared to the profile of a star in the same field. We chose the star 0949-0186220 \citep[NOMAD1,][]{zac04} of magnitude R\,$=$\,14.95 to convert the GRB afterglow intensity profile into a series of magnitudes. Images obtained after the trailed one were performed with a classical sidereal tracking, and co-added to increase the signal to noise ratio for more accurate measures. Like the other optical data, the magnitudes were finally converted into the AB system. These data are listed in Table~\ref{table_data}, and presented as red squares in Fig.~\ref{fig_lc_all}.

\begin{table}
	\centering
	\caption{Zadko and VIRT data. The magnitudes are given in AB system.}
	\label{table_data}
	\begin{tabular}{ccccc}
	\hline\hline
	Start (s) & End (s) & R mag & 2$\cdot\sigma$ error & Telescope \\
	\hline
	59 & 65 & 18.06 & 0.70 & Zadko \\
	65 & 71 & 18.61 & 0.80 & Zadko \\
	71 & 77 & 18.64 & 0.80 & Zadko \\
	77 & 83 & 17.93 & 0.61 & Zadko \\
	83 & 89 & 17.31 & 0.35 & Zadko \\
	89 & 95 & 17.09 & 0.28 & Zadko \\
	95 & 101 & 16.83 & 0.22 & Zadko \\
	101 & 107 & 16.94 & 0.25 & Zadko \\
	107 & 113 & 16.86 & 0.23 & Zadko \\
	113 & 119 & 16.72 & 0.20 & Zadko \\
	137 & 167 & 16.31 & 0.01 & Zadko \\
	177 & 207 & 16.20 & 0.01 & Zadko \\
	216 & 246 & 16.23 & 0.01 & Zadko \\
	255 & 285 & 16.22 & 0.01 & Zadko \\
	295 & 325 & 16.30 & 0.01 & Zadko \\
	334 & 424 & 16.26 & 0.01 & Zadko \\
	534 & 720 & 16.73 & 0.01 & Zadko \\
	832 & 1219 & 17.11 & 0.01 & Zadko \\
	1231 & 1952 & 17.65 & 0.02 & Zadko \\
	2152 & 3000 & 18.16 & 0.04 & Zadko \\
	3079 & 4263 & 18.41 & 0.06 & Zadko \\
	4328 & 5327 & 18.83 & 0.05 & Zadko \\
	5338 & 6777 & 19.03 & 0.12 & Zadko \\
	31690 & 35093 & 19.7 & 0.3 & VIRT \\
	36172 & 40003 & 20.3 & 0.3 & VIRT \\
	40011 & 54065 & 21.0 & 0.4 & VIRT \\
	70413 & 82773 & 21.6 & 0.4 & Zadko \\
	\hline
	\end{tabular}	
\end{table}

\subsection{The VIRT data}

Eleven images of 300\,s with 1\,$\times$\,1 binning were acquired in the clear filter by the VIRT starting from Feb.\,$3^{rd}$\,2017 03:16 UTC. We registered and combined these images to increase the signal to noise ratio. The same reference star from the Zadko images was then used to calibrate the flux density of the afterglow (i.e., 0949-0186220, NOMAD1) in the VIRT data for a more precise correlation between the two instruments. By comparing our reference star on all the VIRT images, we ensured said change in behavior of the GRB afterglow was real, and not due to mis-calibration. The importance of this procedure to our work will become clear as the concatenated light curves of Zadko and VIRT present evidence to intensity variations not expected by the fireball model (see $\S$~\ref{fireball}). The VIRT data are also presented in Fig. \ref{fig_lc_all} as red squares.

As indicated in $\S$~\ref{telescopes}, a second night of GRB 170202A observations was performed on Feb.\,$4^{th}$\,2017 by VIRT. Again, the data were lost during Hurricane Irma that struck the US Virgin Islands in September 2017. 

\section{Fireball modeling}
\label{fireball}

\subsection{Extraction of the temporal decays}

While the fireball model does not expect spectral features on top of the powerlaw segments, the situation is very different with actual light curves. There, late internal shocks, refreshed shocks, and fluctuations of the surrounding density can provide significant variability that can complicate temporal analysis. The lower panel of Fig. \ref{fig_lc_all} presents the multi-spectral light curves and decay indices obtained using a spline fit to our data for GRB 170202A. The high degree of fluctuation in the decay indices is a result of keeping the smoothness parameter low to allow the fit to match the maximum number of data points. This easily facilitates the detection of any correlated fluctuation between the optical and X-ray, and indicates deviations from the standard model that can be filtered out. 

We defined a flare as an episode of rising flux followed by a steep decay seen in optical and/or X-ray. Those not compatible within 3$\sigma$ of a simple power law from the light curves were removed, and a final fit performed using only the power law segments expected by the standard model.

There is a small X-ray flare at $\sim$\,20\,ks post burst. The optical data around this time confirm its optical presence. A second, but poorly sampled, flare could also be present in X-ray at $\sim$\,190\,ks. Our optical data does not cover this epoch. Its last data points, however, present an inflection in the decay indices, which could be suggestive of an achromatic flare (see Fig.~\ref{fig_lc_all}). 

Optical flares in GRB afterglows have been studied by \citet{laz02}, who demonstrated that they do not interfere with the global dynamics of the fireball. We therefore ignored them to study the global evolution of the fireball parameters, but do discuss them later after describing the underlying global emission. 

We point out that no obvious change of behavior was witnessed in the optical light curve during and after the X-ray plateau (Fig.~\ref{fig_lc_all}), and that it would have been possible to group the optical phases "X-ray plateau" and "normal decay" (see Table \ref{table_fit_temporal}) for the analysis into a single phase. Nonetheless, we chose to split the analysis by taking into account the two phases observed in X-ray. 

With the above filtering completed, we extracted decay indices for multiple segments in the X-ray and optical light curves. Table~\ref{table_fit_temporal} summarizes these results along with the temporal definitions we defined for each segment. 

\begin{table}
	\centering
	\caption{List of temporal indices ($\alpha$) of the light curves in Fig.~\ref{fig_lc_all}. Note that we use the standard convention $F \propto t^{-\alpha}$, so a negative $\alpha$ indicates an increase of the light curve.}
	\label{table_fit_temporal}
	\begin{tabular}{lccccc}
	\hline\hline
	Band & Segment           & Start (s) & End (s) & $\alpha$ \\
	\hline
	X-ray   & Steep Decay    & ---    & 400    & $3.1 \pm 0.2$\\
	        & Plateau phase  & 400    & 5000   & $-0.2 \pm 0.1$\\
	        & normal decay   & 5000   & 270000 & $1.1 \pm 0.1$\\
	        & late decay     & 270000 & ---    & $2.3 \pm 0.3$\\
	\hline
	Optical & Rising part    & ---    & 150    & $-3 \pm 1$\\
	        & Slow rising    & 150    & 400    & $-0.2 \pm 0.2$\\
	        & X-ray plateau  & 400    & 5000   & $0.9 \pm 0.2$\\
	        & Normal decay   & 5000   & ---    & $0.8 \pm 0.2$\\
	\hline
	\end{tabular}
\end{table}

\subsection{The steep decay and the duration of the event}
According to \citet{zha06}, the initial part of the X-ray light curve in most GRBs relates to the prompt phase. In this model, the initial steep decay observed in X-ray is the high latitude emission of the prompt phase, i.e., a delayed flux of photons from the end of the prompt phase \citep{kum00}. This led \citet{str13} to define T$_X$, the temporal break at the start of the steep decay, as the true end time of emission from the central engine, and the best estimate possible for the duration of the phenomenon.

Following \citet{kum00}, we should have a closure relation between the spectral and decay indices during the steep decay phase, with $C_1 = \alpha - \beta = 2$. As noted in Tables \ref{table_specX} and \ref{table_fit_temporal}, we measured $\alpha = 3.1 \pm 0.2$ for the steep decay phase of the X-ray light curve and $\beta = 0.94 \pm 0.09$ for its spectral index. This leads to $C_1 = 2.2 \pm 0.3$. The closure relation is thereby valid, and we deduce T$_X \leq 100$\,s.

Before 100\,s the optical data hint to a small flare. We recognize that the large errors and lack of sampling at early times accompanying these data make it difficult to validate this event. It is only mentioned here  because this kind of flaring activity has been seen since the beginning of GRB observations \citep{ake00}, with a short duration that does not influence subsequent emission.

One might question why the optical rising index is so poorly constrained, when comparing to the X-ray emission. This is because the fit is dependent of the smoothness of the peak within our model.

After 100\,s the prompt emission of GRB 170202A has ceased. Therefore, the large rise seen in the optical data between $\sim$ 80 and $\sim$ 200 s should not be linked with the prompt phase, but rather with one of the two remaining components of the fireball: the forward shock and the reverse shock.

\subsection{Plateau phase and reverse shock}

The \citet{wil07} two-component model of X-ray afterglows attributes the plateau phase as the afterglow onset. Coherently, the start time of the X-ray plateau phase is indeed coincident with the start of the decay of the optical afterglow (Fig.~\ref{fig_lc_all}). However, prior to the X-ray plateau, the optical brightness increases with $\alpha = -3$ (see Table~\ref{table_fit_temporal}). The steepest rise expected from the afterglow model is $-2$ \citep{pan00}, which is barely compatible with our observed value. If we consider that the faintest optical data point is due only to the afterglow, and we assume its maximal rise rate from Table~\ref{table_fit_temporal}, then all data earlier than 225\,s are located above the theoretical expectation. This indicates an excess of emission over that time. On the other hand, if we assume the slowest possible rise, an excess of emission exists until 2160\,s. As a consequence we assume that data prior to 225\,s are due to a reverse shock, and data located between 225\,s to 2160\,s possibly relate to this phenomena. It is difficult to be more precise. Usually, the reverse shock emission should not smoothly merge into the forward shock emission, but rather appears as a break within the light curve \citep{zha18}. We can note (at around 330\,s) a small dip into the light curve, which could be interpreted as the end of the reverse shock occurring a little earlier than the peak of the forward shock. However, this dip being so poorly significant (about 2 sigma), we avoid speculating further about the real date of the end of the reverse shock.

\subsection{Electron distribution parameter and surrounding medium density law within the forward shock}

Using the late afterglow behavior of GRB 170202A, attributable to a forward shock, we derived the model parameters via the methods of \citet{gen07}. We started with the properties of the surrounding medium, and the electron distribution parameter, $p$.

Using Table 4 of \citet{gen07} and the X-ray spectral and temporal indices of GRB 170202A, we can assert that the fireball is expanding in the slow cooling mode, with a cooling frequency ($\nu_c$) located below the X-ray band ($\nu_X$). The optical decay index, however, is not the same as the X-ray one. Instead it is compatible with a break of $\Delta\alpha = 0.25$. This indicates the cooling frequency is located above the optical band ($\nu_{opt}$), and the injection frequency ($\nu_i$) is below this band. Such a configuration, and value of $\Delta\alpha$ implies that the fireball is expanding into an interstellar medium (ISM) of (more or less) constant density. 

The near constant ISM density case corresponds to the equations of Appendix B in \citet{pan00}, which we use to evaluate $p$ with our spectral and temporal indices. We have 4 indirect and independent measurements of $p$: 3 decay indices (2 in X-ray, one in optical), and one spectral index. We obtained $p = 2.3 \pm 0.3$, $p = 2.1 \pm 0.2$, $p = 1.9 \pm 0.2$, and $p = 2.1 \pm 0.3$. Reconstructing from these values the probability distribution value of $p$, and deriving its $90\%$ confidence interval, we obtained $p = 2.05 \pm 0.05$. This value will be used throughout the remainder of this paper. 

The optical data between the afterglow peak and the late flares exhibit no temporal break, i.e., an absence of any specific frequency crossing. We turn this into the following condition,
\begin{equation}
\label{eq_ordre}
    \nu_i < \nu_{opt} < \nu_c < \nu_X,
\end{equation}
for GRB 170202A.

\subsection{Spectral Energy Distribution and position of the injection frequency}

\begin{figure*}
	\centering
		\includegraphics[width=8cm]{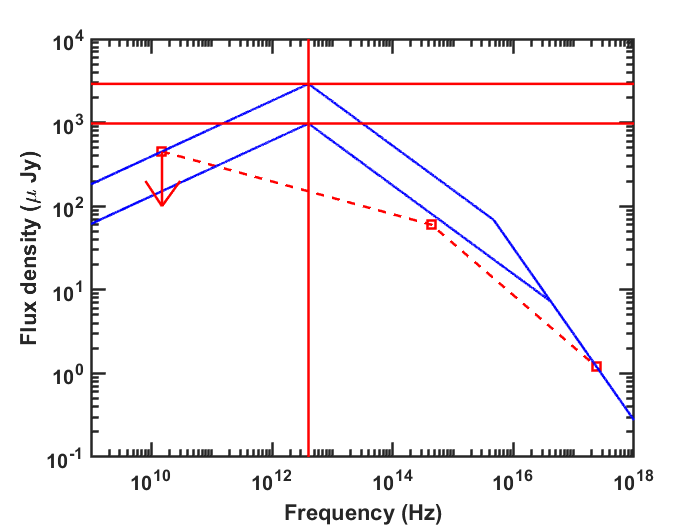}
		\includegraphics[width=8cm]{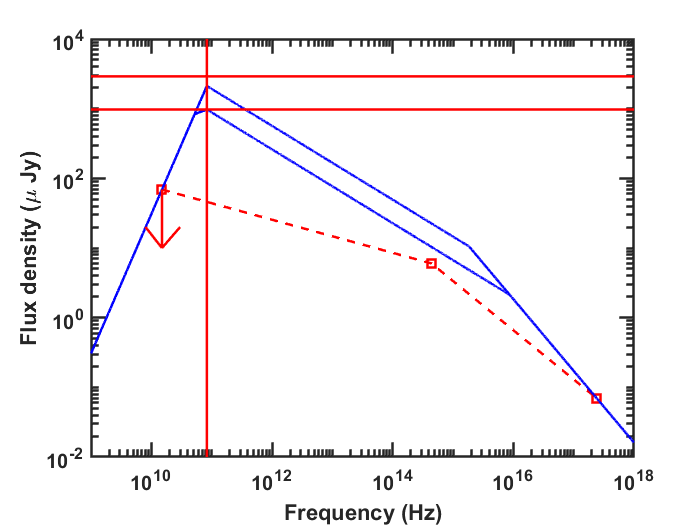}
		\caption{SED of the GRB 170202A fireball at the time of the first (left panel) and second (right panel) radio observations (see Fig.~\ref{fig_lc_all}). We indicate in red the constraints imposed by the model (see text for the explanations) and in blue its limits. \label{fig_SED}}
\end{figure*}

For a shock propagating within the ISM, the peak of each spectral band relates to the passage of the injection frequency and a constant (unabsorbed) flux. We start with the optical band, whose peak at $400 \pm 30$\,s corresponds to the start of the decay phase of the optical afterglow, as stated previously. 

In a pure forward shock model, before this epoch, the decay index should be $-0.5$, but  we observe $-0.2 \pm 0.2$. While this is not in agreement with the standard model, we already highlighted that a portion of our observed emission likely relates to a reverse shock. This phenomenon may explain the difference in the decay index. Consider the model of GRB 170202A developed so far, where a reverse shock could exist until $\sim$\,2000\,s post trigger with contributions to observed emission not bright enough to mask the peak of the afterglow. To compute the position of its injection frequency, we need to have some information about $\epsilon_{e, -1}$ and $\epsilon_{B, -2}$. We can use fact that at the peak of the emission of the forward shock in optical, by definition we are observing at the injection frequency. Doing so, we have:
\begin{equation}
\label{eq_nu_i}
    \nu_i = 2.19 \times 10^{17} \epsilon_{e, -1}^2 \epsilon_{B, -2}^{1/2} = 4.4 \times 10^{14} \mbox{ Hz}.
\end{equation}

It is now possible to compute the position of the injection frequency at any time, and we do so at the time of the radio observations. This is denoted by the red vertical line in Fig.~\ref{fig_SED}. The flux at this time is 974 $\mu$J. Note that this gives the maximum of the Spectral Energy Distribution (SED) at any time. Within Fig.~\ref{fig_SED}, we denote this by  
the lower red horizontal line, which corresponds to a model with no extinction.

The cooling frequency provides another constraint on the SED, as we know its position is above the optical band and below the X-ray band at any time (see Equation~\ref{eq_ordre}). This constraint decays with time, so we use the earliest X-ray and latest optical measurements from the period of normal decay. This X-ray measurement results in, 
\begin{equation}
\label{eq_nu_c}
        \nu_c = 1.45 \times 10^{12} n_{0}^{-1.02} \epsilon_{e, -1}^{-1.07} \epsilon_{B, -3}^{-0.53} < 2.4 \times 10^{17} \mbox{ Hz}.
\end{equation}
The optical value sets a maximum flux at $\nu_i$ of 2.9 mJy. This is denoted in Fig.~\ref{fig_SED} by the upper red horizontal line. As this result was constructed with X-ray data, where the host absorption is negligible, our model corresponds with the maximal amount of extinction due to the host. The final model therefore must fall between these constraints to fit the global observations, taking into account the real host extinction. 

As these constraints evolve with time, we refined them by considering the upper limits for GRB 170202A reported in radio. To do so, the first step is to construct the SED at the time of each radio observation, and then to compute the evolution of the constraints. Fig.~\ref{fig_SED} presents the results with the radio observation at $t = $\,8685\,s (left), and the one at $t = $\,106317\,s (right).

Interestingly, we find that the second radio upper limit, which is deeper than the first one, implies that the self-absorption frequency is located at $\nu_a >$\,15\,GHz. This does not aid in constraining the first SED, so it was left out of Fig.~\ref{fig_SED} (left panel) for clarity. Our upper radio limit does force a lower limit on $\nu_a$, namely, 
\begin{equation}
\label{eq_nu_a}
    \nu_a = 2.6 \times 10^9 E_{53}^{1/5}n_0^{3/5}\epsilon_{e,-1}^{-1}\epsilon_{B,-2}^{1/5} > 15 \times 10^9 \mbox{Hz}.
\end{equation}

\subsection{Optical extinction and surrounding density}

The value of the surrounding density parameter is usually obtained through radio measurements \citep[e.g.][]{gen07}. However, in the case of GRB 170202A, no radio detection was reported (see Section~\ref{telescopes}). This required an {\it ad hoc} fix. In the right panel of Fig.~\ref{fig_SED}, it is seen that the optical data requires an amount of extinction linked to the host galaxy, to match the expected model. We derived a value of A$_R$ between 0.58\,--\,1.43. 

We cannot discern the type of extinction law that should be used (i.e., Milky Way vs. Magellanic Cloud laws), but we can better constrain the value of the density parameter. Our 0.58\,--\,1.43 range of optical extinction is typical of a normal ISM density. This rules out the case of either $n_0$\,$>>$\,1, or $n_0$\,$<<$\,1. It is worth while to point out, the effect of this parameter on our previous equations scales with a power far less than one. Therefore, its effect on the outcomes of this study are small, and we safely assume that $n_0$\,$\sim$\,1 is a sufficient first-order approximation to the surrounding medium density of GRB 170202A.

\subsection{The microphysics parameters}

Returning to our equations for specific frequencies, we can now derive the mircrophysics parameters of the GRB 170202A fireball. Specifically $\epsilon_e$ and $\epsilon_B$ that give the fireball energy distribution between electrons and the magnetic field, respectively. Using Eq.~\ref{eq_nu_i} we obtain,
\begin{equation}
\label{eq_micro3}
    \epsilon_{e} = 1.41 \times 10^{-3} \epsilon_{B}^{-1/4},
\end{equation}
and from B8 of \citet{pan00} the flux observed in X-ray during the second SED leads to, 
\begin{equation}\label{eq_microFnu}
    \begin{split}
        F_\nu &= 3.40 \times 10^{-4} \epsilon_{e, -1}^{1.05} \epsilon_{B, -2}^{0.01}, \\ &= 0.7 \times 10^{-4}\mbox{mJy}.
    \end{split}
\end{equation}
By combing Eqs.~\ref{eq_micro3} and \ref{eq_microFnu} we then arrive at,
\begin{equation}
\label{eq_micro4}
    \begin{split}
        \epsilon_{e} &= 0.0216,\\
        \epsilon_{B} &= 1.84 \times 10^{-5}. 
    \end{split}
\end{equation}

\subsection{The jet and its opening angle}

The late break observed in the X-ray light curve could be due to either a jet break, or a cooling break (Fig.~\ref{fig_lc_all}). As argued below, the former hypothesis is favored by three tests we can apply to our data. 

Firstly, a cooling break would imply a difference on the decay indices of $\Delta \alpha = 0.25$ \citep{pan00}. This is significantly less than the $\Delta \alpha = 1.2 \pm 0.3$ we found for GRB 170202A. Secondly, our value of $\Delta \alpha$ would imply an unrealistic value of the electron distribution index of the burst, i.e., $p \sim 3.7$. Lastly, a jet break should be achromatic, while a cooling break should not \citep{pan00,kum00}. We lack optical data necessary to confirm achromatic behavior at this period in our burst, and its optical limits offer no further constraints to our model. Despite the inconclusive nature of this last test, the first two strongly oppose the null hypothesis. Thus, we consider the late break of GRB 170202A results from the effects of a relativistic jet. 

With the surrounding medium structure determined, we derived the jet aperture angle. For the ISM case, we use the work of \citet{sar99b} to write,
\begin{equation}
    \label{eq_angle}
    \theta = 0.002 E_{52}^{-1/8} t_b^{3/8} n_0^{1/8} = 0.13 n_0^{1/8},
\end{equation}
whose result is expressed in radians. In Eq.~\ref{eq_angle}, note that $t_b$ is the jet break time in seconds, and the X-ray data of GRB 170202A gives $t_b = 270000$ s (see Table~\ref{table_fit_temporal}). Using these results we arrive at $\theta \sim 7.5$ \degree, which is typical for GRBs \citep{fra01,fon15}.

\subsection{The deceleration radius}

The deceleration radius is reached when the afterglow emission stops brightening in all wavelengths. Because the initial afterglow emission is likely mixed with that of a reverse shock and late prompt emission, a precise time cannot be extracted for its deceleration. Though we can consider that the deceleration of this burst started when the optical emission started to decay, i.e., $t_{dec} < 400 \pm 30$\,s (Fig.~\ref{fig_lc_all} and Table~\ref{table_fit_temporal}). This value constrains the initial Lorentz factor of the fireball according to the equation,
\begin{equation}
    \label{eq_deceleration}
    \Gamma_{2.47} = 1.15 t_{dec, rest}^{-3/8} E_{52}^{1/8} n_0^{-1/8} > 1.02 n_0^{-1/8}.
\end{equation}
Plugging in $n_0 = 1$ (as discussed previously), we find $\Gamma > 300$ for GRB 17020A.

\begin{table*}
\centering
\caption{Parameters of the GRB 170202A fireball model derived from observations used in this study. The values of GRB 110205A have been shown for comparison purposes only in the right part of the table.}
\label{table_model}
\begin{tabular}{ccc|cc}
\hline\hline
Parameter & Hypothesis value & Model value & Hypothesis value & Model value\\
\hline
Radiative efficiency (\%)  &  30 &                   --- &   30 &   --- \\
E$_0$ (ergs)               & --- & $56.7 \times 10^{52}$ &  --- & $145 \times 10^{52}$  \\ 
$n_0$                      & 1.0 & ---                   &  0.1 &   --- \\
$p$                        & --- & 2.05                  &  --- &   2.2 \\ 
$\epsilon_e$ forward shock & --- & 0.02                  &  --- &  0.01 \\
$\epsilon_B$ forward shock & --- &  $1.8 \times 10^{-5}$ &  --- & 0.008 \\
$\Theta_j$ (\degree)       & --- & 7.5                   &  --- &   2.1 \\
$\Gamma$                   & --- & $>$300                &  --- &   125 \\

\hline
\end{tabular}
\end{table*}

\subsection{The flares}

We finally focus on the last component of our model, the flares. There are two episodes of flare-like rebrightenings for GRB 170202A (Fig.~\ref{fig_lc_all}). The first visible both in optical and X-ray and peaking at about 20\,ks post-trigger. A second rebrightening was only visible in X-ray at $\sim$\,190\,ks, but as we already pointed out optical data was not available for this epoch, so further discussions are avoided. 

Note that the optical component of the first flaring event is brighter than that of the X-ray. This contradicts ``normal" X-ray flares witnessed in other bursts, i.e., the optical component is dimmer \citep[see][for one example]{kru09}. From the model we built so far, this type of flaring event is not possible from either density fluctuations or the injection of energy. In fact, its X-ray flux should be independent of the ambient density because the X-ray frequency is above the cooling frequency \citep{kum00b}. Alternatively, in the case of a late injection of energy, we should expect a correlated and scaled behavior in the X-ray and the optical light curves. This is not what has been observed: the optical flare is brighter than the X-ray counterpart.

To explain the GRB 170202A flaring event, we return to the argument of a reverse shock. The X-ray component of this flare is considered to follow the widely accepted idea of an injection of energy from a late internal shock \citep[e.g., ][]{zha18}. We then argue that an excess of observed optical emission, compared to its X-ray counterpart, is a combined result of this late internal shock and a reverse shock, which it can easily produce \citep[e.g., see][ and our discussion below]{gao13}. 

\citet{gao15} suggested that the magnetic energy in a reverse shock is roughly 100 times greater than that of the forward shock. For GRB 170202A, the late mixing of the forward shock, refreshed energy injection, and flaring with contributions from a late reverse shock make it impossible to disentangle the energetic details of such a reverse shock. With that said, the value $\epsilon_B$ within the reverse shock can not be quantified, but a value of $\sim$\,100\,$\times$\,$\epsilon_B$ (of the forward shock), i.e. $2 \times 10^{-3}$ would reasonably explain the optical excesses observed for the late GRB 170202A flare.

\section{Discussion and conclusion}
\label{discu}

Table~\ref{table_model} summarizes our model of GRB 170202A. The bulk of it was constrained by observations, with exception of the surrounding medium density and the position of the peak of the afterglow emission. We assumed that the peak of the afterglow corresponded to the last observed optical point before the onset of a decay in its light curve. While this assertion does influence the value of $\Gamma$, it has no effect on $\epsilon_e$ and $\epsilon_B$. For comparison, consider the values of $\epsilon_e$ and $\epsilon_B$ that were derived similarly for GRB 110205A, e.g., see Table~\ref{table_model} and \citet{gen12}. 

The Lorentz factor of GRB 170202A is far greater than that of GRB 110205A. GRB 110205A was known to be a ``slow" event, in that it allowed each component of the fireball to be observed one right after the other. GRB 170202A seems to be a more common burst, where all the fireball components are observed more or less simultaneously. Given the compliance of GRB 170202A with the Amati relation (see Section \ref{telescopes}), we consider that the fireball parameters of this burst are representative of a typical GRB.  

The $\epsilon_e$ and $\epsilon_B$ results for GRB 170202A are far smaller than those expected by the standard fireball model, i.e, the $\epsilon_e \sim 0.1$ and $\epsilon_B \sim 0.01$ of \cite{pan00}. Our results, however, are in agreement with the work of \citet{gao15}, who reported a similar partition between the electron and magnetic field energy using Monte-Carlo simulations. We also point out that the $\epsilon_B$ reported for GRB 110205A is far larger than found for GRB 170202A, while the opposite can be said of their initial Lorentz factors (see Table~\ref{table_model}). This observation is interesting as it may suggest a link exists between the initial Lorentz factor and the magnetic energy of the fireball. 

The complexity from mixing afterglow components could have been misconstrued at the onset of this study as an inability to model the fireball of GRB 170202A sufficiently. We have shown such is not the case, and that pending the availability of a comprehensive enough data-set, it is feasible to extract the relevant physics. Unfortunately, the reverse shock parameters are not well constrained for GRB 170202A. Such could be obtained for future bursts if the temporal coverage were more dense than those available to this study. This emphasizes the importance of and need for a global network of telescopes providing continuous observations of GRBs for several days. 

Our study has exemplified this conclusion by its use of observations from the Zadko and VIRT telescopes. The location of these two telescopes in Western Australia and the Caribbean, respectively, places them at roughly antipodes of each other, i.e., one at night while the other is at day, where coordinated observations allow for complete coverage of any source visible in the nocturnal sky. Fortunately, both followed GRB 170202A until its optical emission completely faded, as it was this extended sampling which uncovered the late optical flare in its afterglow.

An attempt to reproduce our analysis without knowledge of the late GRB 170202A optical flare was carried out. The resultant optical decay index was incompatible with either the X-ray data or the standard fireball model. We found this result interesting, particularly as it is the common approach of literature to dismiss such discrepancies as being explained by late flaring activity, which in turn leads to an inability to derive all the parameters of the fireball model. A main takeaway of our study then, is that it highlights the utility of airing on the side of caution when attempting to explain discrepancies to the fireball model while relying on a light curve with {\it scarce} sampling. 

Dedicated instruments like the ZTF \citep{bel19} and emerging networks like the Global Rapid Advanced Network Devoted to Multi-messenger Addicts \cite[GRANDMA,][]{ant20} have been specifically designed for detecting transient events. The size of the field-of-view (or equivalent field-of-view in the case of a network of telescopes) and the observation methods employed by these groups limit their usefulness to studies such as undertaken for GRB 170202A here. Specifically, their observational protocols are not tuned to provide well-sampled (near-continuous) coverage of the initial nor final phases of optical transients. The fast reactivity of the control software used by Zadko allowed the rising component of the optical emission of GRB 170202A to be captured. It was the scientific trade-off relationship to the size of the field-of-view of the Zadko and VIRT telescopes versus the rate of potential new interesting events that encouraged their continual observations of this burst until it completely faded, while simultaneously ignoring any new transient detections. It is not an intent of our study to downplay the importance of instruments and collaborations dedicated to detecting the most transient sources at any give time. We only wish to highlight how our work has shown an equivalent importance exists for extended and coordinated observing campaigns of individual transient events from their detection to complete fading. This conclusion should encourage other telescope teams to increase the observational time they dedicate to the optical counterparts of transient events to aid the scientific community in developing a more complete picture of the nature of GRBs. Moreover, to facilitate such future efforts, we would also encourage detection teams to consider providing a unique flag that indicates a ``burst of interest" for extended follow-up, possibly based off promising results from their initial properties.

\section*{Acknowledgements}

We would like to thank the anonymous referee for a very dedicated report and checks of this work. We gratefully acknowledge support through NASA-EPSCoR grant NNX13AD28A. B.G. and N.B.O. also acknowledge financial support from NASA-MIRO grant NNX15AP95A, and NASA-RID grant NNX16AL44A. P.G. and N.B.O. acknowledge financial support from NSF-EiR grant 1901296. E.M. acknowledges support support from the Zadko Undergrad Fellowship. Parts of this research were conducted by the Australian Research Council Centre of Excellence for Gravitational Wave Discovery (OzGrav), through project number CE170100004. This research has been partly made under the auspices of the FIGARONet collaborative network supported by the Agence Nationale de la Recherche, program ANR-14-CE33. We also dedicate this work to the late Jim Zadko, whose donation of the Zadko Telescope to the University of Western Australia and continuous support of this facility made possible this study.

\end{document}